# Performance Analysis of AODV under Black Hole Attack through Use of OPNET Simulator

H. A. Esmaili
Department of Computer Engineering, Sharif University of Technology, Tehran, Iran
h_esmaili@ce.sharif.edu

M. R. Khalili Shoja
Department of Electrical Engineering, Amirkabir University of Technology, Tehran, Iran
m.khalili@aut.ac.ir

Hossein gharaee
Department of Electrical Engineering, Tarbiat Modares University, Tehran, Iran
gharaee@itrc.ac.ir

*Abstract*— Mobile ad hoc networks (MANETs) are dynamic wireless networks without any infrastructure. These networks are weak against many types of attacks. One of these attacks is the black hole. In this attack, a malicious node advertises itself as having freshest or shortest path to specific node to absorb packets to itself. The effect of black hole attack on ad hoc network using AODV as a routing protocol will be examined in this research. Furthermore, we investigate solution for increasing security in these networks. Simulation results using OPNET simulator depict that packet delivery ratio in the presence of malicious nodes, reduces notably.

## I. INTRODUCTION

Ad-hoc networks are characterized by dynamic topology, self-configuration, self-organization, restricted power, temporary network and lack of infrastructure. Characteristics of these networks lead to using them in disaster recovery operation, smart buildings and military battlefields [1].

Routing protocol in ad-hoc networks are classified into two main categories, proactive and reactive [3]. In proactive routing protocols, routing information of nodes is exchanged, periodically, such as DSDV [4]. In on-demand routing protocols, nodes exchange routing information when needed such as, AODV [2] and DSR [5]. Furthermore, some ad-hoc routing protocols are a combination of above categories.

Although trusted environment has been assumed in most research on ad-hoc routing protocols, many usages of ad-hoc network run in untrusted situations. So, most ad-hoc routing protocols are vulnerable to diverse types of attacks that one of which is black hole attack. In this attack, a malicious node uses the routing protocol to advertise itself as having the shortest or freshest path to the node whose packets it wants to intercept. In a flooding based protocol, the attacker listens to requests for routes. When the attacker receives a request for a route to the target node, the attacker creates a reply consisting of an extremely short or fresh route [6]. The rest of this paper is organized as follows: In section 2, AODV routing protocol is described. In section 3, we describe classification of attacks in MANET. Network layer attack is described in section 4. Section 5 summarizes solutions to achieve secure level in ad hoc networks. In section 6, simulation results are analyzed.

## II. AODV ROUTING PROTOCOL

AODV is used to find a route between source and destination as needed and this routing protocol uses three significant type of messages, route request (RREQ), route reply (RREP) and route error (RERR). Field information of these messages, such as source sequence number, destination sequence number, hop count and etc is explained in detail in [2]. Each node has a routing table, which contains information about the route to the specific destination. When source node wants to communicate with a destination and there is not any route between them in its routing table, at first step source node broadcasts RREQ. So, RREQ is received by intermediate nodes that they are in the transmission range of sender. These nodes broadcast RREQ until RREQ is received by destination or an intermediate node that has fresh enough route to the destination. Then it sends RREP unicastly toward the source. Hence, a route among source and destination is made. A fresh enough route is a valid route entry that its destination sequence number is at least as great as destination sequence number in RREQ. The source sequence number is used to determine freshness about route to the source. In addition, destination sequence number is used to determine freshness of a route to the destination. When intermediate nodes receive RREQ, with consideration of source sequence number and hop count, make or update a reverse route entry in its routing table for that source. Furthermore, when intermediate nodes receive RREP, with consideration of destination sequence number and hop count, make or update a forward route entry in its routing table for that destination.

### III. CLASSIFICATION OF ATTACKS IN MANET

The attacks in MANET can be classified into two categories, called passive attacks and active attacks. Passive attacks are done to steal information of network such as, eavesdropping attacks and traffic analysis attacks. Indeed, passive attackers get data exchanged in the network without disrupting the operation of a network and modification of exchanged data. On the other hand, in active attacks, replication, modification and deletion of exchanged data is done by attackers. The attacks in ad-hoc networks can also be classified into two categories, called external attacks and internal attacks. Internal attacks are done by authorized node in the network, whereas external attacks are executed by nodes that they are not authorized to participate in the network options. Another classification of attacks is related to protocol stacks, for instance, network layer attacks.

### IV. NETWORK LAYER ATTACKS IN MANET

Some network layer attacks are described in below:

*A. Wormhole attack*

In this attack, an attacker records a packet, at one location in the network, tunnels the packet to another location and replays it there [21].

*B. Byzantine attack*

In this attack, malicious nodes individually or cooperatively carry out attacks such as creating routing loops and forwarding packets through non-optimal paths.

*C. Rushing attack*

Rushing attacker forwards packets quickly by skipping some of the routing processes. So, in on-demand routing protocol such as AODV, the route between source and destination include rushing nodes.

*D. Resource consumption attack*

In this attack, an attacker attempts to consume battery life of other nodes.

*E. Location disclosure attack*

In this attack, information relating to structure of network is revealed by attacker nodes.

*F. Black hole attack*

In black hole attack, malicious nodes falsely claim a fresh route to the destination to absorb transmitted data from source to that destination and drop them instead of forwarding.

Black hole attack in AODV protocol can be classified into two categories: black hole attack caused by RREP and black hole attack caused by RREQ.

*1) Black hole attack caused by RREQ*

With sending fake RREQ messages an attacker can form black hole attack as follows:

*a) Set the originator IP address in RREQ to the originating node's IP address.*

*b) Set the destination IP address in RREQ to the destination node's IP address.*

*c) Set the source IP address of IP header to its own IP address.*

*d) Set the destination IP address of IP header to broadcast address.*

*e) Choose high sequence number and low hop count and put them in related fields in RREQ.*

So, false information about source node is inserted to the routing table of nodes that get fake RREQ. Hence, if these nodes want to send data to the source, at first step they send it to the malicious node.

*2) Black hole attack caused by RREP*

With sending fake RREP messages an attacker can form black hole attack. After receiving RREQ from source node, a malicious node can generate black hole attack by sending RREP as follow:

*a) Set the originator IP address in RREP to the originating node's IP address.*

*b) Set the destination IP address in RREP to the destination node's IP address.*

*c) Set the source IP address of IP header to its own IP address.*

*d) Set the destination IP address of IP header to the IP address of node that RREQ has been received from it.*

### V. SOLUTIONS TO ACHIEVE SECURE LEVEL IN AD HOC NETWORKS

There are basically two approaches to secure MANET: 1.securing ad-hoc routing and 2.Intrusion detection [7].

*A. Securing routing*

Ariadne [8] has proposed ad-hoc routing protocol that provides security in MANET and relies on efficient symmetric cryptography. This protocol is based on the basic operation of the DSR protocol. In [9], a secure routing protocol based on DSDV has been proposed. Hash chains have been used to authenticate hop counts and sequence numbers. ARAN [10] uses cryptographic public-key certificates in order to achieve the security goals. The goal of SAR [11] is to characterize and explicitly represent the trust values and trust relationships associated with ad-hoc nodes and use these values to make routing decisions. Secure AODV (SAODV) [12] is a security extension of AODV protocol, based on public key cryptography. Hash chains are used in this protocol to authenticate the hop count. Adaptive SAODV (A-SAODV) [13] has proposed a mechanism based on SAODV for improving the performance of SAODV. In [14] a bit of modification has been applied to A-SAODV for increasing its performance. TRP [20] employs hash chain algorithm to generate a token, which is appended to the data packets to identify the authenticity of the routing packets and to choose correct route for data packets. TRP provides significant reduction in energy consumption and routing packet delay by using hash algorithm.

### B. Intrusion detection system

[15] introduces a method that requires each intermediate node to send back the next hop information inside RREP message. This method uses further request message and further reply message to verify the validity of the route. Zhang and Lee [16] propose a distributed and cooperative intrusion detection model based on statistical anomaly detection techniques. In [17], the intermediate node requests its next hop to send a confirmation message to the source. After receiving both route reply and confirmation message, the source determines the validity of path according to its policy. In [18], Huang et al use both specification-based and statistical-based approaches. They construct an Extended Finite State Automation (EFSA) according to the specification of AODV routing protocol and model normal state and detect attacks with anomaly detection and specification-based detection. An approach based on dynamic training method in which the training data is updated at regular time intervals has been proposed in [19].

TABLE I. SIMULATION PARAMETERS

| Simulation parameters | Value |
|---|---|
| Number of nodes | 46 |
| Network size | 600*600(m) |
| Simulation duration | 600(sec) |
| Transmit power(w) | .0001 |
| Packet Reception-power Threshold(dBm) | -95 |
| Hash function | SHA-1 |
| Source node | Mobile-node-1 |
| Destination node | Mobile-node-4 |
| Packet Inter-Arrival Time(sec) | Uniform(.1,.11) |
| Packet size(bits) | Exponential(1024) |

### VI. SIMULATION RESULTS & CONCLUSION

For the simulation, we use OPNET 14.0.A [22] as a simulator. Our network topology is indicated in Fig. 1. TABLE I contains parameters that we choose for simulation. For evaluating the performance of the network, we consider the following metric:

Packet Delivery Ratio: The ratio of the data delivered to the destination to the data sent out by the source.

Various mobilities of nodes have been considered to measure the performance of network in presence of malicious nodes as attackers. Fig. 2 demonstrates the results in presence of only one malicious node. Results in Fig 3, 4 and 5 have been obtained in existence of 2, 3 and 4 attackers. The average of packet delivery ratio in each case has been shown in table 2. As it is revealed by these figures, PDR reduces remarkably in presence of black hole attackers.

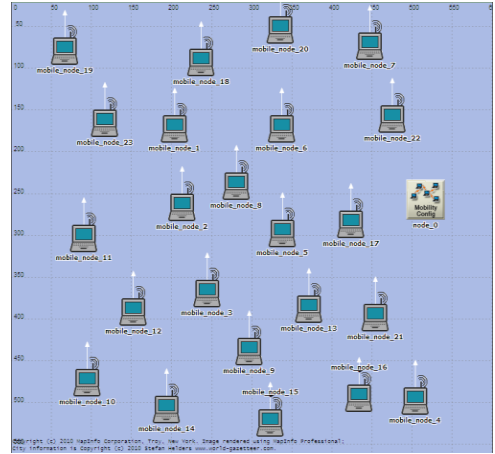

Figure 1. Network topology

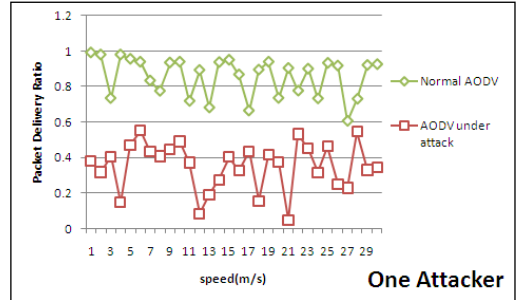

Figure 2. PDR in presence of one malicious node

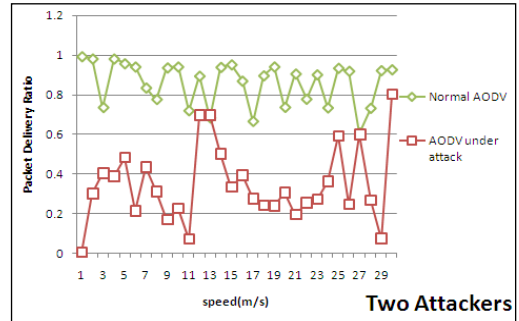

Figure 3. PDR in presence of two malicious node

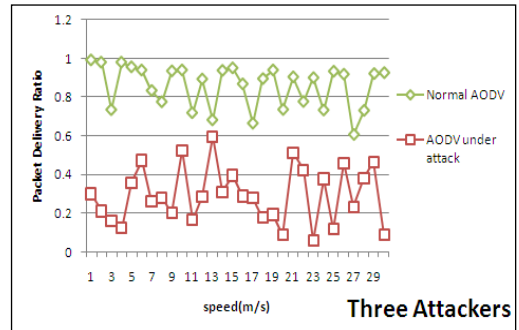

Figure 4. PDR in presence of three malicious node

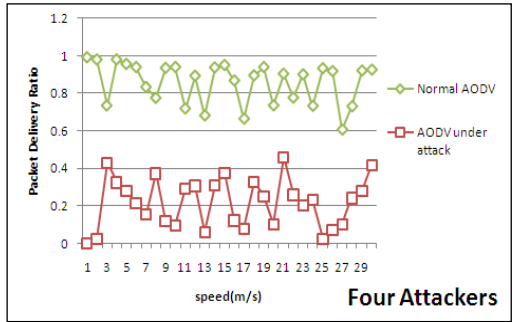

Figure 5. PDR in presence of four malicious node

TABLE II. THE AVERAGE OF PACKET DELIVERY RATIO VS. DIFFERENT NUMBER OF ATTACKERS

|  | Normal AODV | Normal AODV under attack |
|---|---|---|
| One attacker | 0.8578 | 0.3525 |
| Two attackers | 0.8578 | 0.3445 |
| Three attackers | 0.8578 | 0.2934 |
| Four attackers | 0.8578 | 0.2179 |